# Orbital Engineering in Three Dimensional Halide Perovskites towards Two Dimensional Properties with Strong Anisotropy

Gang Tang,[a] Zewen Xiao,[b] Jiawang Hong [a,] *

[a] School of Aerospace Engineering, Beijing Institute of Technology, Beijing, 100081, China

[b] Wuhan National Laboratory for Optoelectronics, Huazhong University of Science and Technology, Wuhan 430074, China

**Corresponding Author**

*E-mail: hongjw@bit.edu.cn.

**ABSTRACT**

The discovery of double perovskites $A_2B(I)B(III)X_6$ (A = monovalent cation; B(I)/B(III) = metal cation; X = halogen) as Pb(II)-free alternatives has attracted widespread attention, making it possible to introduce d-block metal cations (e.g., $d^0$ and $d^{10}$) into halide perovskites. However, at present, there are quite limited insights into the underlying bonding orbitals for d-block metal cation-based halide perovskites. Here, we present an orbital engineering strategy to construct two-dimensional (2D) electronic structures in three-dimensional (3D) halide perovskites by rationally controlling the d orbitals of the metal cations to hybridize with the halide p orbitals. Taking $Cs_2Au(I)Au(III)I_6$ as an example, we demonstrate that the flat conduction band and valence band at the band edges can be achieved simultaneously by combining two metal cations with different d orbital configurations using the first-principles calculations. The predicted carrier mobilities show huge anisotropy along the in-plane and out-of-plane directions in $Cs_2Au(I)Au(III)I_6$, further confirming the 2D electronic properties. In addition, the anisotropic static dielectronic constants and Young's modulus are also observed. More importantly, it is found that $Cs_2Au(I)Au(III)I_6$ has excellent mechanical flexible and an ultra-small shear modulus among halide perovskites. Our work provides valuable guidance for achieving low-dimensional electronic characteristic in three-dimensional halide perovskites for novel electronic applications.

**TOC GRAPHICS**

# 1. Introduction

Orbital engineering plays an important role in understanding and manipulating the various electronic and magnetic behaviors in solids, especially in transition metal oxides (TMO).[1, 2, 3, 4, 5, 6] Generally, the electronic structure of TMO is dominated by the electronically active d orbitals of transition metal (TM) cation and the neighbouring oxygen 2$p$ ligand orbitals.[2] As a result, manipulating the $d$ orbital orders of TM cation offers the opportunities to realize novel physical phenomena and improved properties in various fields such as topological insulators and superconductors.[3, 7] In addition, several theoretical studies have revealed that the anisotropic-shaped d orbitals of TM cation hybridizing with the oxygen p orbitals result in the low-dimensional nature of the lowest conduction bands.[8, 9] As the representative of $d^0$ perovskites, the most commonly used $SrTiO_3$, its two-dimensional character of the conduction bands resulted from the planar character of the (pdπ) interaction[8], and its application in two-dimensional electron gases (2DEGs) has received widespread attention[10]. For example, many experimental and theoretical studies have focused on investigating the 2DEGs at oxide interfaces (e.g., $LaTiO_3/SrTiO_3$)[10, 11, 12], which may exhibit unique properties such as extremely high electron densities, magnetism, and 2D superconductivity. It is also highly desirable to discover similar electronic phenomena in other perovskite systems.

In recent years, lead (Pb)-based halide perovskites $APbX_3$ (A = monovalent cation; X = halogen) have attracted great attention and have emerged as excellent candidates for optoelectronic applications such as solar cells, light-emitting diodes, and photodetectors.[13, 14, 15] Despite the intriguing features of Pb halide perovskites, the presence of toxic Pb in the chemical composition is considered as one of the major factors hindering their commercialization.[16] Some strategies have been suggested to search for Pb-free perovskite alternatives.[17] To date, one of the most promising approaches is that mutating two Pb(II) into a pair of B(I)/B(III) cations or a B(IV) cation to form halide double perovskites $A_2B(I)B(III)X_6$ ( B(I) = Tl$^+$, Ag$^+$, Au$^+$; B(III) = Sb$^{3+}$, Bi$^{3+}$, In$^{3+}$, Au$^{3+}$)[18] or vacancy-ordered defect-variant perovskites $A_2B(IV)X_6$ (B(IV) = Sn$^{4+}$, Pd$^{4+}$, Ti$^{4+}$)[19]. The recently discovered $A_2B(I)B(III)X_6$ and $A_2B(IV)X_6$ series compounds significantly widening the possible combination of metals with different oxidation states to replace Pb. On the other hand, the formation of a double perovskite structure makes it possible to introduce some *d*-block elements (e.g., $d^0$ or $d^{10}$ metal cations) into the halide perovskites, providing us new orbital degrees of freedom to modulate the optoelectronic properties. However, at present, there are very limited insights into the underlying bonding orbitals at the band edges of the *d*-block metal cations-based halide perovskites, let alone manipulating their properties through orbital engineering of *d* electron orbitals.

In this work, we demonstrate that the d/p-hybridization in halide perovskite results in two-dimensional nature of conduction and valence bands at the band edges with different cation *d* orbital filling states. We propose an orbital engineering strategy to construct two-dimensional electronic structures in three-dimensional halide perovskites. Taking mixed-valence double perovskite Cs$_2$Au(I)Au(III)I$_6$ as an example, we demonstrate that the flat conduction band and valence band at the band edges can be achieved by combining two metal cations with different d orbital configurations using the first-principles calculations. The carrier mobilities are calculated to further confirm the 2D electronic properties. In addition, the static dielectronic constants and mechanical properties are also explored in Cs$_2$Au(I)Au(III)I$_6$.

## 2. Results and discussions

**Orbital characteristics of *d*-band halide perovskites.** Figure 1a-c illustrates the schematic of orbital engineering to construct two-dimensional electronic structures in the three-dimensional halide perovskites. According to previous reports[8, 9, 20], the *d/p*-hybridization tends to create the low-dimensional electronic structures, which is reflected in two-dimensional dispersion at the band edges. The occurrence of two-dimensional dispersion at the conduction band edge or the valence band edge will depend on the degree of the cation *d* orbital filling. If the cation d orbital is empty (fully filled), the metal-halide orbital interactions will result in the occurrence of two-dimensional dispersion at the conduction band (valence band) edge, as shown in the Figure 1a and 1b. Therefore, it is natural to think about whether two-dimensional dispersion at the conduction band and valence band edge can occur simultaneously by combining two different d metal cations (see Figure 1c).

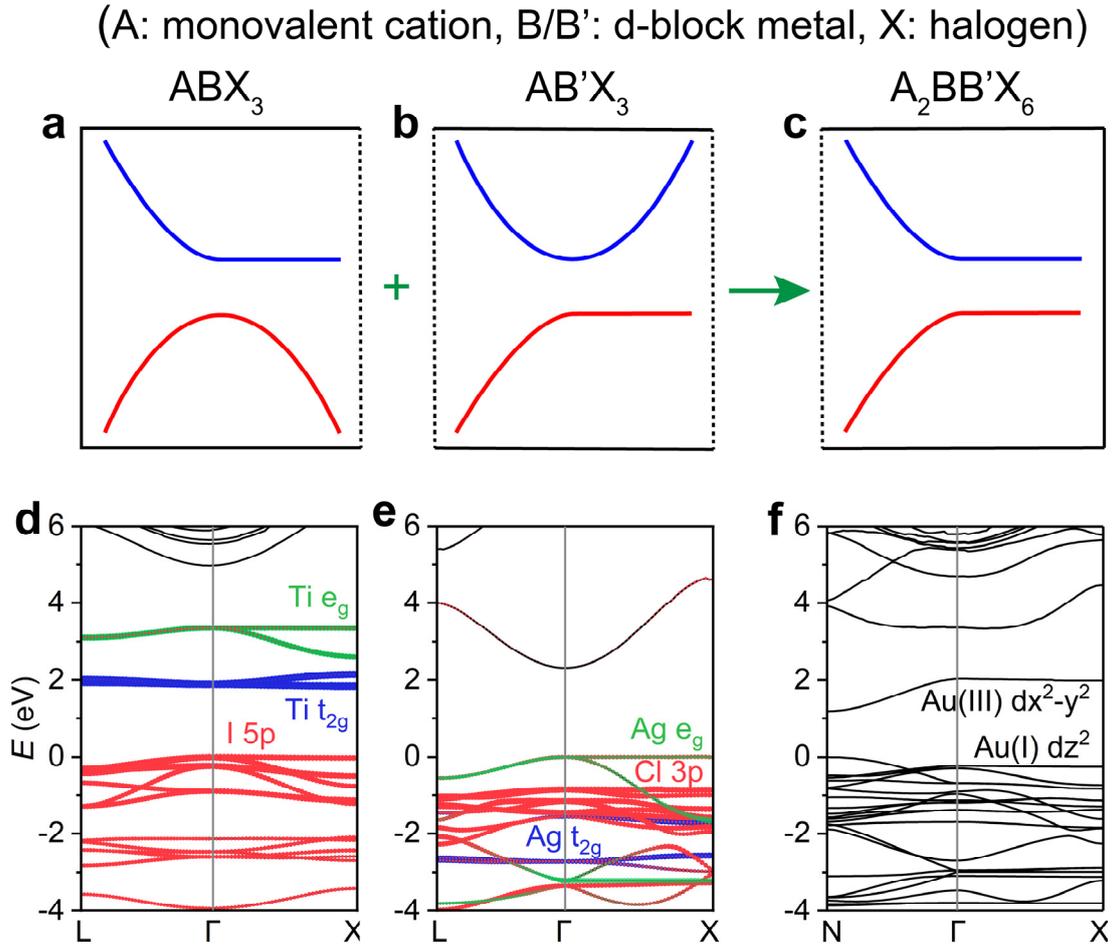

**Figure 1.** Orbital engineering to realize two-dimensional character of the conduction bands and valence bands in three-dimensional halide perovskites. (a-c) shows the schematic of orbital engineering for *d*-block metal cations-based halide perovskites. (a)/(b) shows the B/B' *d*-X *p* hybridization creates the two-dimensional character of the valence/conduction bands. (c) shows that two-dimensional band edges are achieved by combining the two different d-block metals. In order to validate our idea, we have calculated the band structures of three specific materials such as (d) $Cs_2TiI_6$, (e) $Cs_2AgInCl_6$, and (f) $Cs_2Au(I)Au(III)I_6$ corresponding to the three cases in the (a-c).

In order to verify the feasibility of our ideas, we take specific materials as a representative example to analyze the corresponding electronic structures. As the representative of $d^0$ perovskites, we use $Cs_2TiI_6$ as a case study and calculated its band structure, as shown in Figure 1d. $Cs_2TiI_6$ crystallizes in the cubic crystal structure with space group *Fm-3m* (see Figure S1a), which is recently proposed as the nontoxic and stable Pb-free derivatives exhibiting promising optoelectronic properties.[21] As expected, the conduction band edge of $Cs_2TiI_6$ exhibits nondispersion along the Γ-X direction, in well agreement with recent work[21]. This flat band originates from the hybridization of I 5p states and Ti $t_{2g}$ states (see Figure 1d and Figure S2a), leading to a two-dimensional nature of conduction band. $Cs_2AgInI_6$ is chosen as the representative of $d^{10}$ perovskites (see Figure S1b), and the calculated band structure is shown in Figure 1e. Indeed, the flat band between Γ and X occurs at the valence band edge, mainly resulting from the hybridization of Cl 3p states and Ag $e_g$ states (see Figure 1e and Figure S2a). Our findings are in good consistent with previous theoretical studies[22, 23].

In the oxide double perovskites $A_2^{2+}B'B''O_6$, the combination of the two metal cations $B'$ and $B''$ may be $B'^{4+}/B''^{4+}$, $B'^{3+}/B''^{5+}$, $B'^{2+}/B''^{6+}$ or $B'^{1+}/B''^{7+}$,[24] and there are many TM cations including d-orbital available. However, $B'^{1+}/B''^{3+}$ is the only combination in the existed three-dimensional halide double perovskites $A_2^{1+}B'B''X_6$. Therefore, to combine two d-block metal elements with the different oxidation states in the same perovskite structure while satisfying the charge neutrality, such an example is rare in halide perovskite family. Fortunately, the mixed-valence halide double perovskite family $A_2Au(I)Au(III)X_6$[25, 26] satisfy these two criteria. Very recently, Debbichi et al.[27] have proposed that $Cs_2Au(I)Au(III)I_6$ will be a promising absorber for high-efficiency Pb-free thin-film solar cells by means of multiscale theoretical simulations. Therefore, we choose $Cs_2Au(I)Au(III)I_6$ as an example and calculate its band structure, as shown in Figure 1f. It can be seen that the flat conduction band and valence band at the band edges are achieved simultaneously along the Γ-X direction, which results in the two-dimensional electronic properties in a pure three-dimensional halide perovskite. In the next section, we will analyse in detail the origin of the resulting flat bands and its interesting properties.

**Two-dimensional electronic properties of Cs$_2$Au(I)Au(III)I$_6$.** The property of the material is highly related to the corresponding crystal structure, so we firstly focus on the structure of Cs$_2$Au(I)Au(III)I$_6$. Figure 2a shows the tetragonal crystal structure of Cs$_2$Au(I)Au(III)I$_6$ with space group *I*4/*mmm*[28]. As previously reported[25, 26, 28], the Au(I) atom is linearly coordinated by two I atoms, and the Au(III) atom is surrounded by four I atoms in a perfectly square-planar arrangement. Both Au(I) and Au(III) are surrounded by six I atoms when the second nearest neighbours are included. Consequently, its structure consists of three-dimensional (3D) Au-I frameworks formed by the alternating elongated [Au$^{III}$I$_6$] octahedra and compressed [Au$^{I}$I$_6$] octahedra sharing their corners. The elongated and compressed octahedral crystal fields lead to different 5d-orbital splitting of Au$^{III}$ ($d^8$) and Au$^{I}$ ($d^{10}$)[29, 30], as shown in Figure 2b. It is worth noting that the highest energy $5d_{x^2-y^2}$ orbitals of Au(III) are unfilled. As pointed out by Kitagawa et al.[25, 26], from another structural point of view, an anisotropy is present in the 3D -Au$^I$-I-Au$^{III}$-I- networks (see Figure 2a). For example, the networks along the *c* direction are ···I-Au$^I$-I···Au$^{III}$···, while ···Au$^I$···I-Au$^{III}$-I··· along the *ab* plane. Moreover, the Au$^I$···I distance (3.297 Å) in the networks of *ab* plane is notably shorter than the Au$^{III}$···I (3.655 Å) distance in the *c* direction. This observed structural anisotropy is closely correlated to the anisotropic optoelectronic properties, as discussed below.

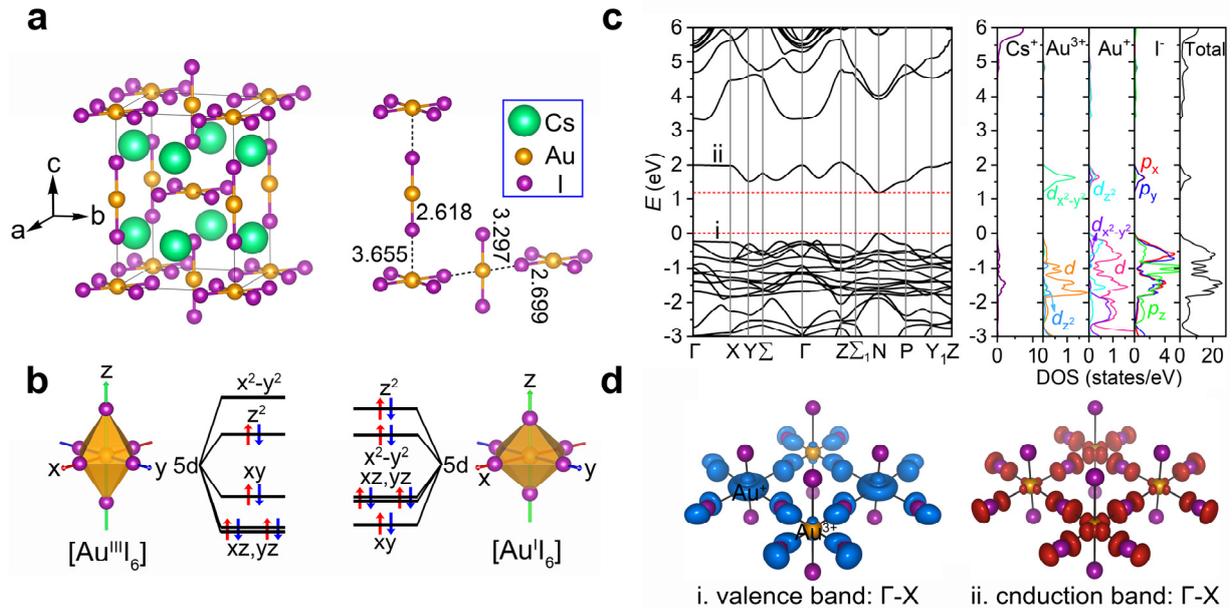

**Figure 2.** Crystal structure, electronic properties, and square modules of the wave functions for the mixed-valence double perovskite $Cs_2Au(I)Au(III)I_6$. (a) Linear $[Au^{I}I_2]^-$ and square-planar $[Au^{III}I_4]^-$ complexes are formed alternately, and right panel shows anisotropic Au-I chains, e.g., ···I-$Au^I$-I···$Au^{III}$··· in the *c* direction and ···$Au^I$···I-$Au^{III}$-I··· in the *ab* plane. Note that the unit for the bond length is Å. (b) Schematic energy level diagram for 5*d*-orbital splitting of $Au^{III}$ ($d^8$) and $Au^I$ ($d^{10}$) under elongated and compressed $[AuI_6]$ octahedral crystal field, respectively. (c) Band structure and projected density of states using the HSE06 method. The zero of the energy scale is set to the top of the valence band. (d) Isosurface plots of the square modules of the Kohn-Sham wave functions corresponding to the top of the valence band along the Γ-X direction and the bottom of the conduction band along the Γ-X direction.

In Figure 2c, we show the electronic properties of $Cs_2Au(I)Au(III)I_6$ from the HSE06 method. It should be noted that, for reasonable band structure calculations, the *k* path should be carefully chosen so that it includes all important directions for carrier transport. The calculated band structure along the high symmetry directions is shown in Figure 2c and the corresponding first Brillouin zone in Figure S3. It can be seen that the valence band maximum (VBM) and the conduction band minimum (CBM) are both located at the N point (0.0, 0.5, 0.0). Therefore, the HSE06 method predicts a direct band gap of 1.19 eV for $Cs_2Au(I)Au(III)I_6$, which is in good agreement with recent theoretical report (1.21 eV)[27] and close to previous experimental result (1.31 eV)[30]. Unlike previously reported Ag(I)-based analogues (e.g., $E_g$ > 1.9 eV)[31], Au(I)-Au(III)-based iodine double perovskite shows a smaller band gap suitable as solar cell absorbers. Such a small band gap is due to the formation of an isolated intermediate band consisting of the lower conduction band, which is attributed to the hybridization between the unoccupied Au(III) $5d_{x^2-y^2}$ orbitals with I $5p_x$ and $p_y$ orbitals coupled with small amount of occupied Au(I) $5d_{z^2}$ orbitals (see Figure 2c). And the VBM is dominated by the Au(I) $5d$ orbitals and I $p$ orbitals with a small contribution from Au(III) $5d$ states.

In addition, the band structure displays significantly dispersive characters along the band gap edges (e.g., $\Sigma_1$-N and N-P directions). The estimated effective masses for holes and electrons along the $\Sigma_1$-N (N-P) directions are 0.34 (0.65) $m_0$ and 0.40 (0.81) $m_0$ ($m_0$ is the electron static mass), respectively. However, there is almost no dispersion for the lower conduction band and the upper valence band along the Γ-X direction (i.e., [001]). Note that previous reports[27, 32] did not contain this important Γ-X path in the calculated band structure. The holes and electrons effective masses along the Γ-X direction are calculated to be 25.94 $m_0$ and 60.16 $m_0$, respectively, indicating that the carriers are difficult to move along the [001] direction. The I $5p$ states are significantly off the centre of the lines of Au-I networks (see Figure 2d), weakening the coupling between the I $5p$ states and Au(I) $5d_{z^2}$ states, which is responsible for the observed flat valence band. This flat conduction band originates from the hybridization of I $5p_{x,y}$ states and $5d_{x^2-y^2}$ states of Au(III) and Au(I) (see Figure 2d), leading to a two-dimensional wave function confined within the equatorial (001) plane.

The dispersion of the band edges directly determines the carrier effective masses and also strongly affects the carrier mobility. Thus we also provide theoretical predictions for the carrier mobilities of $Cs_2Au(I)Au(III)I_6$ along different directions by applying the deformation potential theory as discussed in the Supporting Information. The calculated results are summarized in Table 1 and Figure S4. It can be seen that the carrier mobilities show very strong anisotropic along the in-plane and out-of-plane directions in $Cs_2Au(I)Au(III)I_6$. For example, the predicted electron and hole mobilities in (100) direction are 376.75 and 182.12 $cm^2$ $V^{-1}$ $s^{-1}$. However, the carrier mobilities in (001) direction are quite small (< 0.002 $cm^2$ $V^{-1}$ $s^{-1}$), as reflected by the flat conduction band and valence band edges along the Γ-X direction in Figure 2c. Such huge anisotropy that the mobilities of in-plane are around $10^7$ times larger than that of out-of-plane is very uncommon in a three-dimensional material. We also compare our results with the theoretical intrinsic mobility of prototypical tetragonal phase $MAPbI_3$[33], in which the mobilities in the *ab* plane are only ~4 times larger than that of along the *c* direction (see Table 1). At the same time, it is worth noting that in-plane electron mobility of $Cs_2Au(I)Au(III)I_6$ is smaller compared to the theoretically reported value of $Cs_2AgInCl_6$ double perovskite[34], but the in-plane hole mobility is larger that of $Cs_2AgInCl_6$.

**Table 1.** Predicted carrier mobility of $Cs_2Au(I)Au(III)I_6$. The results of prototypical $Cs_2AgInCl_6$ and $MAPbI_3$ from literatures are also given for comparison.

| Materials | Direction | Carrier type | $m^*$ | $E_1$ (eV) | $C_{ii}$ (GPa) | $\mu$ ($cm^2$ $V^{-1}$ $s^{-1}$) |
|---|---|---|---|---|---|---|
| $Cs_2Au(I)Au(III)I_6$ | (100)/(010) | e | 0.67 | -2.62 | 15.53 | 376.75 |
| | | h | 0.51 | -5.30 | | 182.12 |
| | (001) | e | 107.17 | -4.12 | 17.43 | 0.0005 |
| | | h | 59.39 | -5.08 | | 0.0015 |
| $Cs_2AgInCl_6$ (Ref.[34]) | | e | 0.27 | -10.19 | | 976.79 |
| | | h | 1.01 | -11.05 | | 34.41 |
| CH$_3$NH$_3$PbI$_3$ (Ref.[33]) (Tetragonal) | (100) | e | 0.24 | -1.48 | 23.3 | 23.4 |
| | | h | 0.31 | -3.26 | | 2.50 |
| | (001) | e | 0.21 | -4.40 | 42.3 | 6.70 |
| | | h | 0.28 | -6.41 | | 1.50 |

Note that: carrier types 'e' and 'h' denote 'electron' and 'hole', respectively. $m^*$ represents carrier effective masses. $E_1$ and $C_{ii}$ are the deformation potential and elastic modulus. Mobilities $\mu$ were calculated with the temperature $T$ at 300 K.

**Anisotropic dielectric and optical properties.** Dielectric constants of semiconductor materials are an important descriptor for optoelectronic applications.[35] Large dielectric constant across many orders of magnitude in frequency is beneficial for the charge carrier screening, as revealed in prototype CH$_3$NH$_3$PbI$_3$ by theoretical and experimental studies.[36, 37] In the case of Cs$_2$Au(I)Au(III)I$_6$, the calculated static dielectric tensors are $\varepsilon_{std}^{xx} = \varepsilon_{std}^{yy} = 24.26$ and $\varepsilon_{std}^{zz} = 12.94$ (see Table S1). It can be seen that the values of the dielectric constant along the *ab* plane ($\varepsilon_{std}^{xx}$ and $\varepsilon_{std}^{yy}$) are very close to the theoretically reported value ($\varepsilon_{std} = 25.7$) of CH$_3$NH$_3$PbI$_3$[37]. In addition, the dielectric tensors show strong anisotropy related to the tetragonal symmetry. Specifically, the value of the dielectric constant along the *c* direction ($\varepsilon_{std}^{zz}$) is much smaller than that of in the *ab* plane, indicating the anisotropic screening ability for the charged impurities and defects in Cs$_2$Au(I)Au(III)I$_6$. The anisotropy of dielectric constants is mainly due to the fact that the electron contribution part has a large difference (~ 64%) along the in-plane and out-of-plane directions (See Table S1).

The optical absorption in the visible light region is another critical parameter for assessing the performance of optoelectronic materials.[38] Figure 3a gives the calculated optical absorption spectra of Cs$_2$Au(I)Au(III)I$_6$ for light polarized parallel and perpendicular to the *c* axis, exhibiting a strongly anisotropic character. It can be found that the absorption contribution along the *c* direction ($α_∥$) is much weaker than that in the *ab* plane ($α_⊥$). In particular, there is almost no significantly absorption below 1.5 eV along the [001] direction. These observations are well consistent with previous experimental reflectivity spectra measurements at 300 K[30] and recent theoretical studies[27, 32]. Besides, the absorptions along the *ab* plane ($α_⊥$) show a sharp peak at 1.25 eV, in agreement with the calculated band gap. Previous studies[29] have revealed that the observed peaks (below ~3eV) in the absorption spectra are assigned to intermetal 5d-5d transitions from the [Au$^I$I$_2$]$^-$ to [Au$^{III}$I$_4$]$^-$ complexes. Moreover, it is worth noting that the intermetal 5d-5d transitions also enable a high absorption coefficient exceeding $10^5$ cm$^{-1}$, which is usually observed in the semiconductors with p-p transitions (i.e., MAPbI$_3$, Sb$_2$Se$_3$)[38, 39]. We noted that the previous reported CuTaS$_3$ semiconductor with d-d transitions also exhibits a high absorption coefficient (>$10^5$ cm$^{-1}$).[40]

The pronounced anisotropy of optical absorption between light polarization along the *c* axis and in the *ab* plane in Cs$_2$Au(I)Au(III)I$_6$ leads to a strong linear dichroism for materials in the visible light region.[41, 42] Figure 3b shows the calculated linear dichroism $\triangle κ = κ_⊥ - κ_∥$, along with the linear birefringence $\triangle n = n_⊥ - n_∥$. The large positive value of $\triangle κ$ can be observed for photon energy at 1.4 ~ 2.4 eV, reflecting the strong anisotropic absorption in the energy range. The linear birefringence $\triangle n$, which is the refractive indices for light polarization parallel and perpendicular to the *c* axis, reaches 0.201 at about 890 nm. This value is comparable to the well-known strongly birefringent materials like black phosphorus and TiO$_2$ (see Table S2)[43, 44]. Thus Cs$_2$Au(I)Au(III)I$_6$ has the potential to be used as a polarizer, filter, and waveplate in photonic devices and circuits.

Considering the anisotropic optical properties as well as the electronic structure and dielectric constants discussed above, to maximize the optoelectronic properties, the [001] plane of the grown film of $Cs_2Au(I)Au(III)I_6$ should be paralleled to the substrate in order to facilitate efficient charge transport and optical absorption, similar to the layered Ruddlesden-Popper (RP) perovskite absorbers.

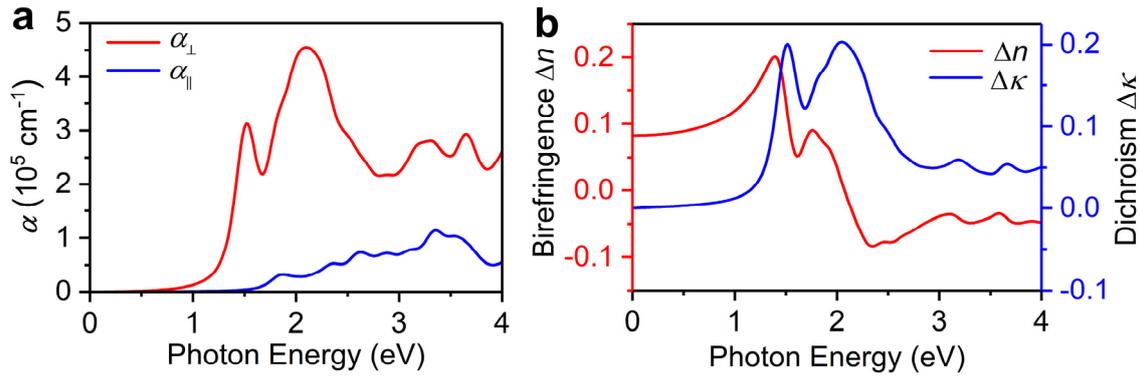

**Figure 3.** Anisotropic optical properties of $Cs_2Au(I)Au(III)I_6$. (a) Optical absorption spectra for light polarization parallel and perpendicular to the $c$ axis (b) Birefringence ($\triangle n = n_\perp - n_\parallel$) and linear dichroism ($\triangle \kappa = \kappa_\perp - \kappa_\parallel$).

**Anisotropic mechanical properties.** Mechanical properties are significantly important for the practical preparation of perovskite solar cells.[45] The elastic constants ($C_{ij}$) of $Cs_2Au(I)Au(III)I_6$ were calculated, as shown in Figure 4a and Table S3. For the tetragonal symmetry, six independent elastic constants can be obtained. We mainly focus on the elastic shear constant $C_{44}$ ($C_{55} = C_{44}$), and its calculated value (0.89 GPa) is at least two times smaller than those of $ABX_3$ and $A_2BB'X_6$ halide perovskites (see Figure 4a). Such low values of $C_{44}$ and $C_{55}$ indicate a quite weak resistance to shear in the (100) and (010) plane, which is unusual in a three-dimensional material. We attribute the ultralow shear modulus to the weak chemical bonding along the $c$ direction because the $Au^{III}\cdots$ I (3.655 Å) distance in the $c$ direction is significantly longer than the $Au^{I}\cdots$I (3.297 Å) distance of the *ab* plane. Based on the obtained elastic constants, the bulk modulus (*B*), shear modulus (*G*), and Young's modulus (*E*) (see Table S4) are also calculated by using the Voigt–Reuss–Hill (*VRH*) approximation. The values of *B*, *G*, and *E* are 8.27, 2.80, and 7.54 GPa in $Cs_2Au(I)Au(III)I_6$, respectively, which are significantly smaller than other halide double perovskites (see Table S4). In addition, the value of *B*/*G* known as Pugh's ratio[46] is greater than 1.75, indicating that $Cs_2Au(I)Au(III)I_6$ has good ductile and is mechanically flexible. Finally, we plotted the directional Young's modulus of $Cs_2Au(I)Au(III)I_6$, as displayed in Figure 4b. It can be seen that $Cs_2Au(I)Au(III)I_6$ exhibits strong anisotropic mechanical properties and has the potential to be used in the field of special flexible electronic devices.

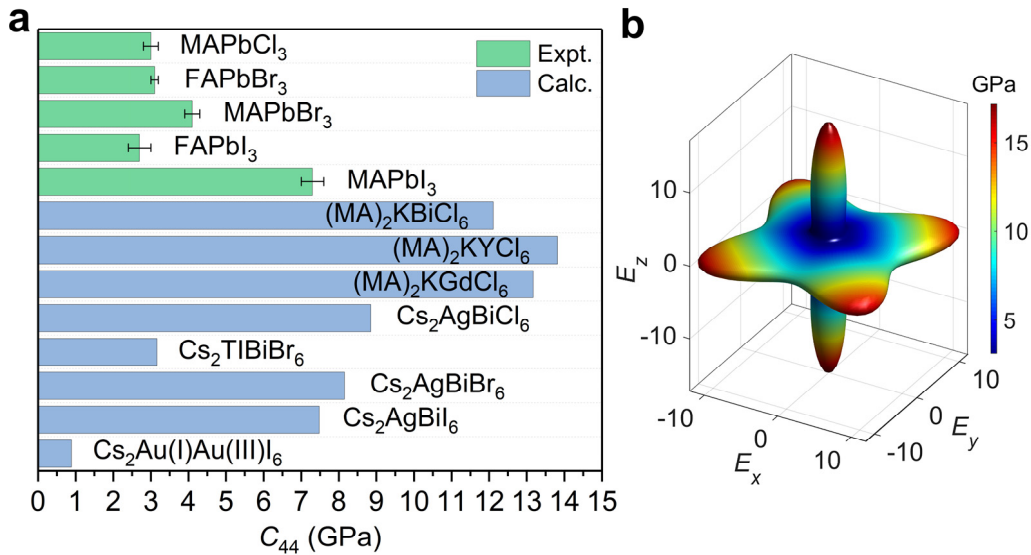

**Figure 4.** Anisotropic mechanical properties. (a) Summary of the calculated $C_{44}$ (GPa) values of halide double perovskites, along with the experimental $C_{44}$ values of traditional halide single perovskites for comparison. The detailed data are given in Table S3. (b) The calculated directional Young's modulus of $Cs_2Au(I)Au(III)I_6$.

**Defect properties.** In addition to the above properties, defect properties of a semiconductor are also quite important for its practical performance.[39] We have therefore examined a range of intrinsic acceptor and donor vacancy defects ($V_{Cs}$, $V_{Au(I)}$, $V_{Au(III)}$, and $V_I$) in $Cs_2Au(I)Au(III)I_6$. The calculated charge-state transition level diagram is shown in Figure 5. It can be seen that $V_{Au(I)}$ is a shallow acceptor with the (-1/0) transitions at 0.16 eV above the VBM, which results from the antibonding coupling between Au(I) 5d and I 5p orbitals raising the VBM. In contrast, as discussed above, partially unoccupied Au(III) 5d orbitals have smaller contribution to the VBM. Therefore $V_{Au(III)}$ exhibits deep transition levels in the band gap [i.e., ε(-1/0) = 1.11 eV and ε(-3/-2) = 0.98 eV]. $V_{Cs}$ is also a shallow acceptor with the (-1/0) transitions at 0.09 eV above the VBM due to the high ionicity of $Cs^+$. These conclusions are generally similar to the previously reported corresponding defects in $Cs_2AgBiBr_6$[47], $Cs_2AgInBr_6$[48], and $Cs_2TIBiBr_6$[49] double perovskites. In contrast to $Cs_2AgBiBr_9$ and $Cs_2AgInBr_6$, in which $V_{Br}$ is a shallow donor, $V_I$ exhibits bipolar features with deep transition levels [i.e., ε(-1/0) = 1.0 eV and ε(0/+1) = 0.60 eV] in the band gap. These deep defects may act as a trap-assisted recombination center, and appropriate growth conditions should be chosen to avoid them during the actual preparation process.

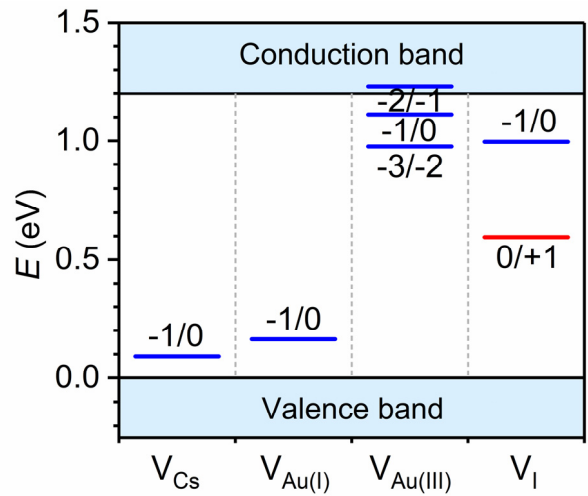

**Figure 5.** Calculated charge-state transition level diagram for a range of intrinsic vacancy in $Cs_2Au(I)Au(III)I_6$.

## 3. Conclusions

In conclusion, we have proposed an orbital engineering strategy to construct two-dimensional (2D) electronic structures in three-dimensional (3D) halide perovskites by rationally controlling the metal cation $d$ orbitals hybridizing with the halide $p$ orbitals. Taking $Cs_2Au(I)Au(III)I_6$ as a case study, we demonstrate that the flat conduction band and valence band at the band edges can be achieved by combining two metal cations with different d orbital configurations using the first-principles calculations. Our calculated carrier mobilities show quite strong anisotropy along the $ab$ plane and $c$ directions in $Cs_2Au(I)Au(III)I_6$, further confirming the 2D electronic properties. In addition, the static dielectronic constants and mechanical properties also display significantly anisotropy, and $Cs_2Au(I)Au(III)I_6$ has better mechanical flexible and a super-small $C_{44}$ values of 0.89 GPa among halide perovskites. Our work provides useful guidance for achieving low-dimensional electronic characteristic in three-dimensional halide perovskites for novel electronic applications.

AUTHOR INFORMATION

**Corresponding Author**
*E-mail: hongjw@bit.edu.cn.
**Notes**
The authors declare no competing financial interests.

ACKNOWLEDGMENT

This work is supported by the National Science Foundation of China (Grant No. 11572040), the Thousand Young Talents Program of China, and Graduate Technological Innovation Project of Beijing Institute of Technology. Theoretical calculations were performed using resources of the National Supercomputer Centre in Guangzhou, which is supported by Special Program for Applied Research on Super Computation of the NSFC-Guangdong Joint Fund (the second phase) under Grant No. U1501501. In addition, Gang Tang thanks Prof. Xing Ming for the insightful discussion.

**Supporting Information Available:**